\documentstyle[prb,aps]{revtex}

\begin{document}

\draft

\title{$^{115}$In-NQR evidence for unconventional superconductivity in CeIn$_3$ under pressure}

\author{S.~Kawasaki$^1$, T.~Mito$^2$, Y.~Kawasaki$^1$, G.-q.~Zheng$^1$, 
Y.~Kitaoka$^1$, H. Shishido$^3$, S.~Araki$^4$, R.~Settai$^3$ and Y.~\=Onuki$^{3,4}$}

\address{$^1$Department of Physical Science, Graduate School of Engineering 
Science, Osaka University,  Osaka 560-8531, Japan}

\address{$^2$Faculty of Science, Kobe University, Kobe, Hyogo 657-8501, Japan}

\address{$^3$Department of Physics, Graduate School of Science, Osaka 
University, Toyonaka, Osaka 560-0043, Japan}

\address{$^4$Advanced Science Research Center, Japan Atomic Energy Research 
Institute, Tokai, Ibaraki 319-1195, Japan}

\date{\today}

\twocolumn[

\maketitle 

\widetext

\begin{abstract}

We report evidence for unconventional superconductivity in CeIn$_3$ at a pressure $P$ = 2.65 GPa above critical pressure ($P_c$ $\sim$ 2.5 GPa) revealed by the measurements of nuclear-spin-lattice-relaxation time ($T_1$) and ac-susceptibility (ac-$\chi$). Both the measurements of $T_1$ and ac-$\chi$ have pointed to a superconducting transition at $T_c$ = 95 mK, which is much lower than an onset temperature $T_{onset}$ = 0.15 K at zero resistance. The temperature dependence of $1/T_1$ shows no coherence peak just below $T_c$, indicative of an unconventional nature for the 
superconductivity induced in CeIn$_3$.
\end{abstract}

\vspace*{5mm}

\pacs{PACS: 74.25.Ha, 74.62.Fj, 74.70.Tx, 75.30.Kz, 76.60.Gv} 

]

\narrowtext
\section{Introduction}
Since the discovery of cerium (Ce) based heavy-fermion (HF) superconductivity in CeCu$_2$Si$_2$ \cite{steglich}, many experimental works on CeCu$_2$Ge$_2$,\cite{CeCu2Ge2} CeIn$_3$ \cite{Mathur,muramatsu,flouquet} and CePd$_2$Si$_2$ \cite{Grosche} have suggested that the antiferromagnetism and superconductivity are closely related each other. The discovery of pressure ($P$) -induced HF superconductivity in Ce-based HF antiferromagnetic (AF) compounds has stimulated further experimental works under $P$.\cite{Hegger,Gegenwart,Kawasaki,Mito} Recently, a new HF material CeRhIn$_5$ consisting of alternating layers of CeIn$_3$ and RhIn$_2$ was found to reveal an AF to superconducting (SC) transition at a relatively lower critical pressure $P_c =$ 1.63 GPa than in previous examples.\cite{CeCu2Ge2,Mathur,Grosche}
The SC transition temperature $T_c$ = 2.2 K is the highest one to date among $P$-induced superconductors.\cite{Hegger}
This finding has opened a way to investigate the $P$-induced evolution of both magnetic and SC properties over a wide $P$ range.
In the previous paper \cite{Mito}, the $^{115}$In-NQR study of CeRhIn$_5$ has clarified the $P$-induced anomalous magnetism and unconventional superconductivity. At $P$ = 2.1 GPa, the nuclear spin-lattice relaxation rate $1/T_1$ reveals a $T^3$ dependence below the SC transition temperature $T_c$, which shows the existence of line-node in the gap function.\cite{Mito} Note that the unconventional superconductivity in CeMIn$_5$($M$ = Co, Rh, Ir) occurs at the border of the AF phase.\cite{Hegger,Mito,Zheng,Shinji}

It is suggested that $P$-induced superconductivity in CeIn$_3$, which is the host crystal of CeMIn$_5$, is also mediated by AF magnetic fluctuations from resistivity measurements under $P$.\cite{Mathur} However, evidence for a bulk superconductivity and whether it is mediated by AF magnetic fluctuations or not in CeIn$_3$ has not been established because its superconductivity under $P$ is observed only in the resistivity 
measurements.\cite{Mathur,muramatsu,flouquet} It is still unclear whether  any problem concerning a sample surface or its quality that could prevent the onset of bulk $P$-induced superconductivity is ruled out or not. This is in contrast to the case of CeRhIn$_5$ where the bulk nature of the SC phase was fully established \cite{Mito,Specificheat} over a wide $P$ range 2 $-$ 5 GPa \cite{muramatsu}. Note that the $P$-induced superconductivity in CeRhIn$_5$ occurs in the non-Fermi-liquid state in the whole $P$ range.\cite{muramatsu}

CeIn$_3$ crystallizes in the cubic AuCu$_3$ structure and orders antiferromagnetically ($T_N = 10$ K) at $P = 0$ with an ordering vector {\bf Q} = (1/2,1/2,1/2).\cite{Morin} $T_N$ monotonically decreases with $P$ and superconductivity appears below $T_c\sim$ 0.2 K at a critical pressure $P_c = 2.55$ GPa and the onset of superconductivity has been observed by the resistivity measurements and is limited in a narrow $P$ range of about 0.5 GPa.\cite{Mathur,Grosche,Walker} Also it is reported that non-Fermi-liquid regime in CeIn$_3$ under $P$ is limited in a quite narrow $P$ range around the peak of resistive $T_c$ in the SC phase.\cite{flouquet} 
The previous $^{115}$In-NQR result on CeIn$_3$ revealed that the AF order disappears around $P^*$ = 2.44 GPa close to $P_c$ and the Fermi-liquid behavior was observed below 10 K at $P = 2.74$ GPa \cite{Kohori-2}.
 In the previous paper \cite{Shinji}, we reported that the localized magnetic character is robust against the application of $P$ up to 1.9 GPa, beyond which $T^{*}$, which marks an evolution into an itinerant magnetic regime, increases rapidly. It is of great interest that the superconductivity in CeIn$_3$ emerges in such an itinerant regime. The window for the 3D AF fluctuation regime with respect to an external pressure is much narrower in CeIn$_3$ than in CeRhIn$_5$, which may also be partly responsible for the narrower SC region in CeIn$_3$.  

In this paper, we report measurements of $1/T_1$ and ac-susceptibility at $P$  = 2.65 GPa down to $T$ = 50 mK which provide evidence for unconventional and bulk nature of superconductivity in CeIn$_3$. At $P$ = 2.65 GPa, from the relation of $T_1T$ = const. below $T_{FL}\sim$ 5K characteristic for the Fermi-liquid state, we conclude that the $P$-induced superconductivity occurs well below $T_{FL}$ without any signature for a non-Fermi-liquid behavior governed by AF spin-fluctuations. Our result is consistent with the recent resistivity measurements.\cite{flouquet} This is in contrast to the case of CeRhIn$_5$ where superconductivity occurs $\it{only}$ in the non-Fermi-liquid regime.\cite{muramatsu}

\section{Experimental}
The single crystals of CeIn$_3$ were grown by Czochralski method as reported 
elsewhere \cite{Onuki} and moderately crushed into grains in order to make rf pulses penetrate into samples easily. To 
avoid some crystal distortions, however, the size of the grains is kept with diameters larger than 100 $\mu$m. 
In a small piece of CeIn$_3$ from the same batch used in this work, the zero resistance transition was confirmed in the 
$P$ range of 2.2 $-$ 2.8 GPa. The phase diagram inferred from the resistivity data is shown in the inset of 
Fig.~4b,\cite{muramatsu} which  is in good agreement with the previous reports.\cite{Mathur,flouquet}  At $P =$ 2.5 GPa 
that is close to a critical pressure $P_c$, the $T_c$ determined by the resistivity measurement reaches a maximum value 
of $T_c$ = 0.19 K and its transition width becomes the sharpest. Measurement of $1/T_1$ was made by the conventional 
saturation-recovery method in the temperature ($T$) range of 0.05 $-$ 100 K at $P$ = 2.65 GPa\@. The $^{115}$In-NQR 
$1/T_1$ was measured at the transition of 3$\nu_{Q}$ ($\pm 5/2\leftrightarrow \pm 7/2$) above $T$ = 1.4 K, but at 
1$\nu_{Q}$ ($\pm 1/2\leftrightarrow \pm 3/2$) below $T$ = 1.4 K. The ac-susceptibility (ac$-\chi$) was measured by using 
an {\it in-situ} NQR coil. Hydrostatic pressure was applied by utilizing a NiCrAl-BeCu piston-cylinder cell, filled with 
Si-based organic liquid as a pressure-transmitting medium. To calibrate the pressure  at low temperatures, the shift in 
$T_{\rm c}$ of Sn metal under pressure was monitored by the resistivity measurement. \cite{smith} To reach the lowest 
temperature of 50 mK, a $^3$He-$^4$He dilution refrigerator is used.

\section{experimental result}
Figures~1 and 2 show the respective $T$ dependencies of $1/T_1$ and $1/T_1T$ at $P =$ 0, 2.35, and 2.65 GPa. At $P$ = 0, 
the AF order occurs at $T_N$ = 10 K without any clear signature for an itinerant HF regime. Rather, in the paramagnetic 
state, the 4$f$-electron derived moments behave as if localized, since $1/T_1$ stays a constant. It was reported that 
$T_N$ monotonically decreases with  pressure and eventually is reduced to $T_N$ = 5 K at $P$ = 2.35 GPa. At this 
pressure, it is noteworthy that $1/T_1$ starts to decrease  below $T^*\sim$ 15 K. This is defined as a characteristic 
temperature below which the system crosses over to the itinerant HF regime.\cite{Shinji} At pressures exceeding $P$ = 2.35 GPa, the internal field at the $^{115}$In site due to the AF ordering decreases, revealing a same $P$ dependence as 
$T_N$. \cite{Kohori-2,thessieu} At $P$ = 2.65 GPa, it is remarkable, as presented in Figs. 1 and 2, that $1/T_1$ shows a 
significant decrease below $T^*\sim$ 30 K, followed by a behavior of $T_1T$ = const. that is characteristic for the 
Fermi-liquid state to be realized below $T_{FL} = 5$ K. This result proves that the Fermi-liquid description is valid 
over a wide $T$ range 0.1 $-$ 5 K at $P$ = 2.65 GPa. Any trace of AF spin-fluctuations at $P$ = 2.65 GPa is not 
appreciable due to an increase of the hybridization of Ce-4$f$ electrons and conduction electrons as discussed in the 
previous paper.\cite{Shinji} This is in good agreement with the recent resistivity measurement. \cite{flouquet} 

Next, we discuss about the superconductivity in CeIn$_3$ at $P$ = 2.65 GPa. 
The SC transition at $P$ = 2.65 GPa is evidenced by the $1/T_1$ measurement. As seen in Figs. 1 and 2, below $T_c$ = 95 
mK, $1/T_1$ decreases suddenly, signaling the onset of bulk SC transition. In addition, the absence of the coherence 
peak in $1/T_1$ just below $T_c$ is consistent with an anisotropic SC gap. The $T_1$ measurement below 50 mK was too 
hard to determine the nodal structure  because of a heat-up problem arising from the exciting rf pulses. 

The SC transition is also corroborated by the ac$-\chi$ measurement. The $T$ dependence of the ac$-\chi$ at $P$ = 2.65 
GPa is shown in Fig.~3\@. The absolute value of susceptibility was determined by using a single crystals of CeIrIn$_5$ 
as a reference in which bulk superconductivity is fully established.\cite{sumiyama} 
The ac$-\chi$ decreases below $T_c\sim$ 95 mK due to the SC diamagnetism, indicating a clear signature of a SC 
transition. The diamagnetic signal of CeIn$_3$ reaches 80$\%$ of the value of CeIrIn$_5$. It indicates the $P$-induced  
superconductivity in CeIn$_3$ is the bulk one.  Note that the $T_c$ probed by the $T_1$ and ac$-\chi$ measurements is 
lower than $T_c^{\rho}$ $\sim$ 0.15 K at zero resistance on the same sample.\cite{muramatsu}  This suggests that the 
surface of the sample makes a short cut for the electrical current flow, exhibiting a rather higher $T_c$ value than 
that for the bulk one.  This phenomenon is  similar to the SC characteristics in CeIrIn$_5$, in which the bulk $T_c 
\sim$ 0.4 K confirmed by the measurements of the specific heat, the susceptibility\cite{Petrovic}, and the NQR 
measurements \cite{Zheng} is much lower than $T_c \sim$ 1 K determined by the zero resistance. A reason why this 
inconsistency occurs is still unresolved.

Note that, at $P$ = 2.65 GPa, the normal state is dominated by the Fermi-liquid excitations below $T_{FL} = 5$ K that is 
far above $T_c = 95$ mK, but not by AF spin-fluctuations. The superconductivity occurs thus in the Fermi-liquid regime, 
contrasting with the unconventional superconductivity in CeCu$_2$Si$_2$\cite{Kawasaki} and CeMIn$_5$\cite{Mito,Zheng} 
both of which reveal a higher $T_c$ near the border of AF and SC transitions in the non-Fermi-liquid state.  Fig. 4a indicates the $T$ dependence of $1/T_1T$ in CeRhIn$_5$ at $P$ = 2.1 GPa where the superconductivity is fully 
established.\cite{Mito} This $T$ dependence was consistent with the AF spin-fluctuation model\cite{Moriya} near a 
magnetic criticality. Most remarkably, the $P$-induced superconductivity in CeRhIn$_5$ occurs under the dominance of 
strong AF spin-fluctuations near the border to the AF phase.  As seen in Figs. 4a and 4b, the behavior of $1/T_1T$ = 
const. in CeIn$_3$ strongly contrasts with that in CeRhIn$_5$. This difference in magnetic fluctuations is suggested to 
be intimately related to the fact that the maximum $T_c^{max} = 2.1$ K in CeRhIn$_5$ is an order magnitude larger than 
$T_c^{max}$ = 0.2 K in CeIn$_3$. Note that $T_c$ is much smaller for CeIn$_3$ because non-Fermi-liquid and AF spin-fluctuations regime is limited in quite narrow $P$ range\cite{flouquet,Shinji} than that in CeCu$_2$Si$_2$ and CeRhIn$_5$. For the latter, the superconductivity occurs at the border of the AF phase where the non-Fermi liquid takes place and even coexists with it. This may be related to the robustness of SC region against increasing $P$ for the latter. \cite{Shinji}  It is an issue to study in future whether or not the superconductivity and the antiferromagnetism in CeIn$_3$ coexist and whether or not the AF magnetic fluctuations in the non-Fermi-liquid state raise $T_c$ near the border of AF and SC phases.
\section{conclusion}
In conclusion, we have provided evidences for the bulk and unconventional superconductivity in CeIn$_3$ below 95 mK at $P$ = 2.65 GPa on the basis of the measurements of $T_1$ and ac-susceptibility. We have found that the system changes the magnetic character from the localized to itinerant regime below $T^*\sim$ 30 K and the Fermi-liquid description becomes valid below $T_{FL}\sim$ 5K without any appreciable trace of AF spin-fluctuations developing. $1/T_1$ reveals no coherence peak just below $T_c$ = 95 mK, indicative of unconventional superconductivity.  We remark that the $P$-induced SC phase in CeIn$_3$ is characterized by a lower $T_c$ and a narrower SC region than in CeRhIn$_5$ where SC phase coexists with the AF phase and $T_c = 2.1$ K is obtained in virtue of the non-Fermi-liquid state dominated by strong AF spin-fluctuations.

\section*{Acknowledgements}
We thank K. Ishida and C. Thessieu for useful discussions.
This work was supported by the COE Research (10CE2004) in Grant-in-Aid for Scientific Research from the Ministry of Education, Sport, Science and Culture of Japan. One of the author ( Yu Kawasaki ) has been supported by JSPS Research Fellowships for Young Scientists.

\begin{figure}[htbp]
\caption[]{$T$ dependence of $ ^{115}(1/T_1)$ in CeIn$_3$ at $P$ = 0, 2.35 and 2.65 GPa. The dotted line indicates a 
relation of $^{115}(1/T_1T)$ = const.. }
\end{figure}

\begin{figure}[htbp]
\caption[]{$T$ dependence of $^{115}(1/T_1T)$ in CeIn$_3$ at $P$ = 0, 2.35 and 2.65 GPa. The dotted line indicates a 
relation of $^{115}(1/T_1T)$ = const..}
\end{figure}

\begin{figure}[htbp]
\caption[]{$T$ dependence of ac-susceptibility in CeIn$_3$ at $P$ = 2.65 GPa. The solid arrow indicates the onset 
temperature of superconductivity.}
\end{figure}

\begin{figure}[htbp]
\caption[]{(a) $T$ dependence of $ ^{115}(1/T_1T)$ in CeRhIn$_5$ at $P$ = 2.1 GPa.\cite{Mito} The solid arrow indicates 
$T_c$. The inset shows the $P$-$T$ phase diagram of CeRhIn$_5$.\cite{Shinji} (b) $T$ dependence of $ ^{115}(1/T_1T)$ in 
CeIn$_3$ at $P$ = 2.65 GPa. The solid (dashed) arrow indicate $T_c$($T_{FL}$). The dotted line indicates a relation of 
$^{115}(1/T_1T)$ = const.. The inset shows the $P$-$T$ phase diagram of CeIn$_3$.\cite{Shinji} }
\end{figure}

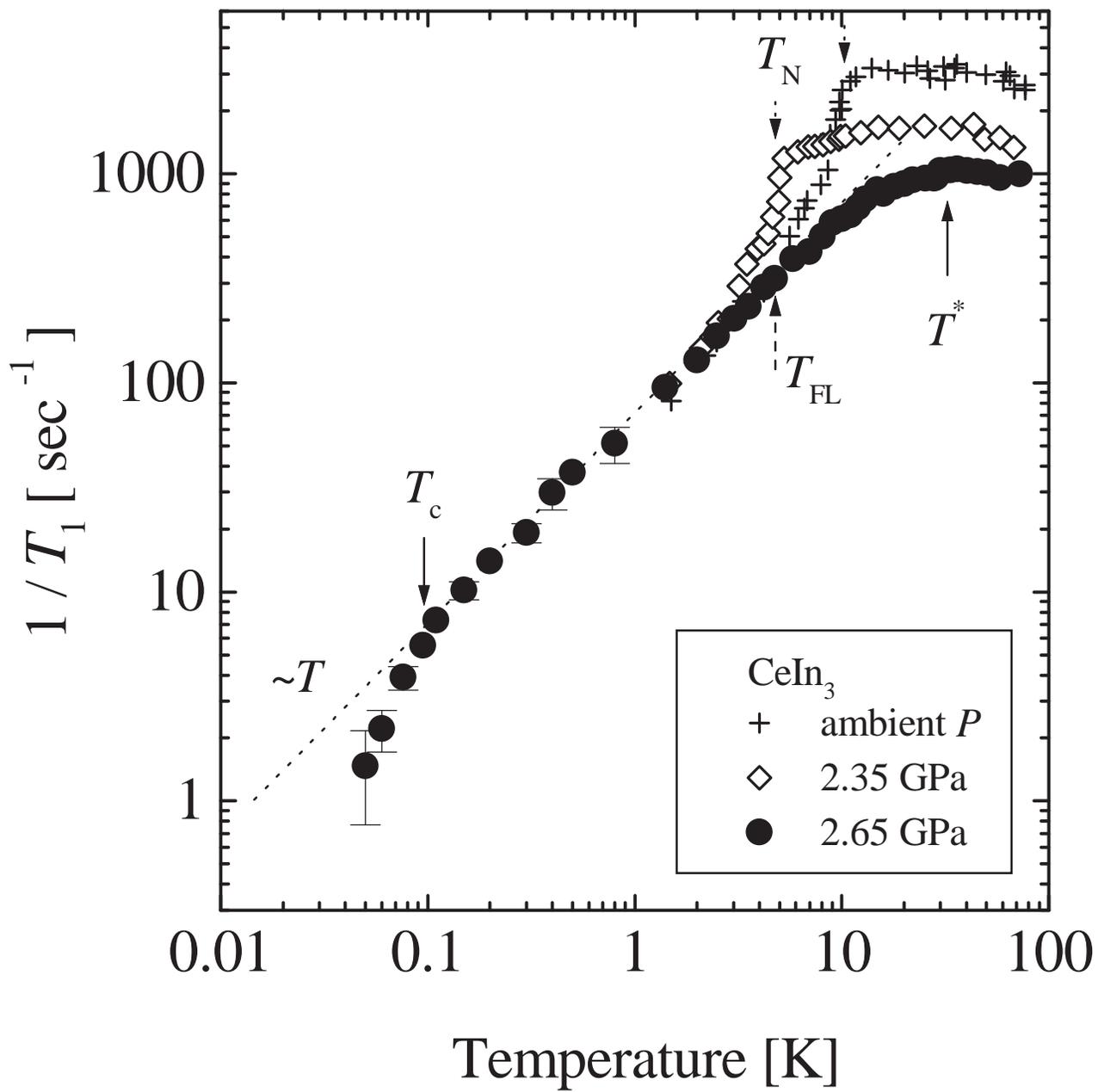
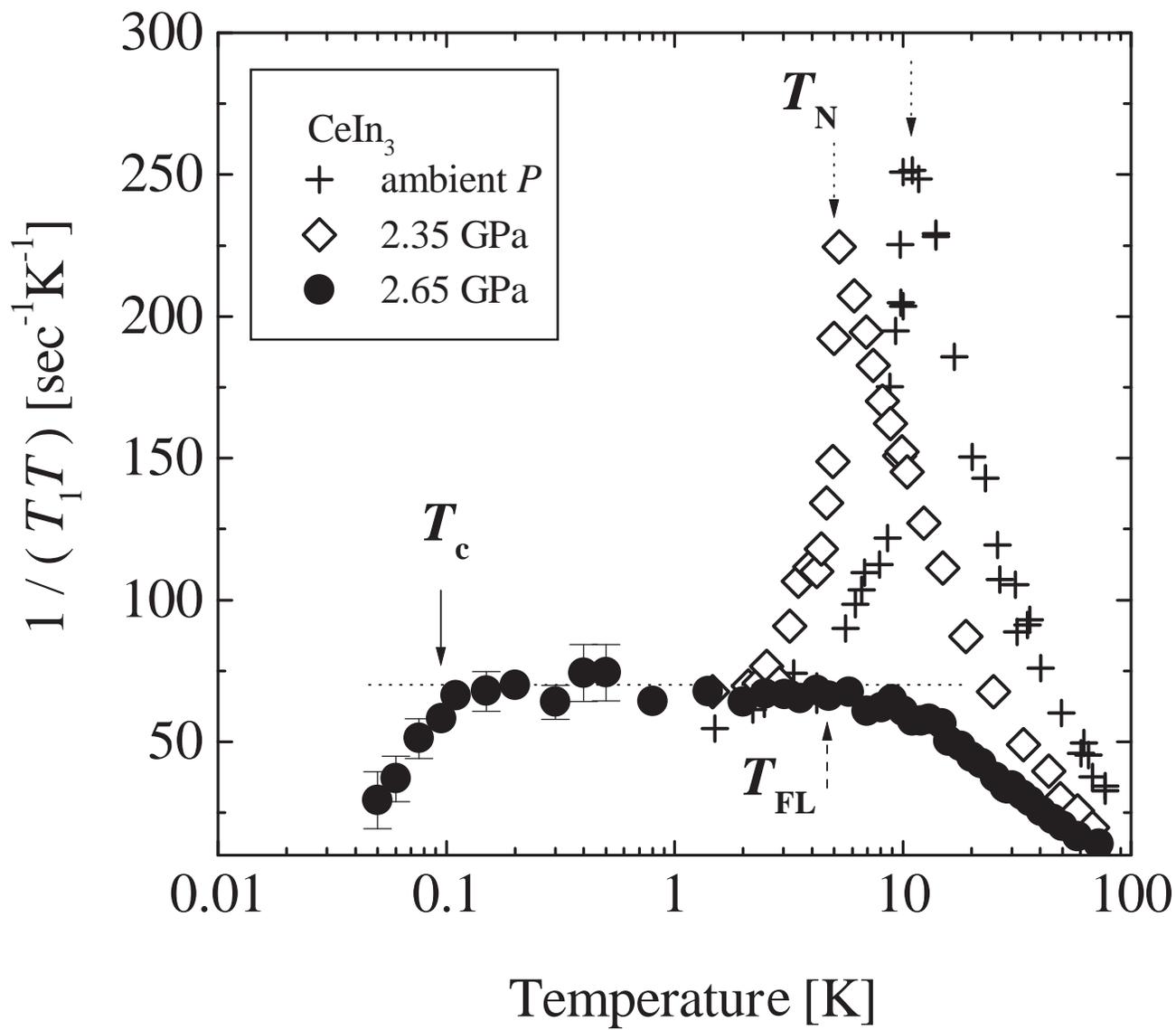
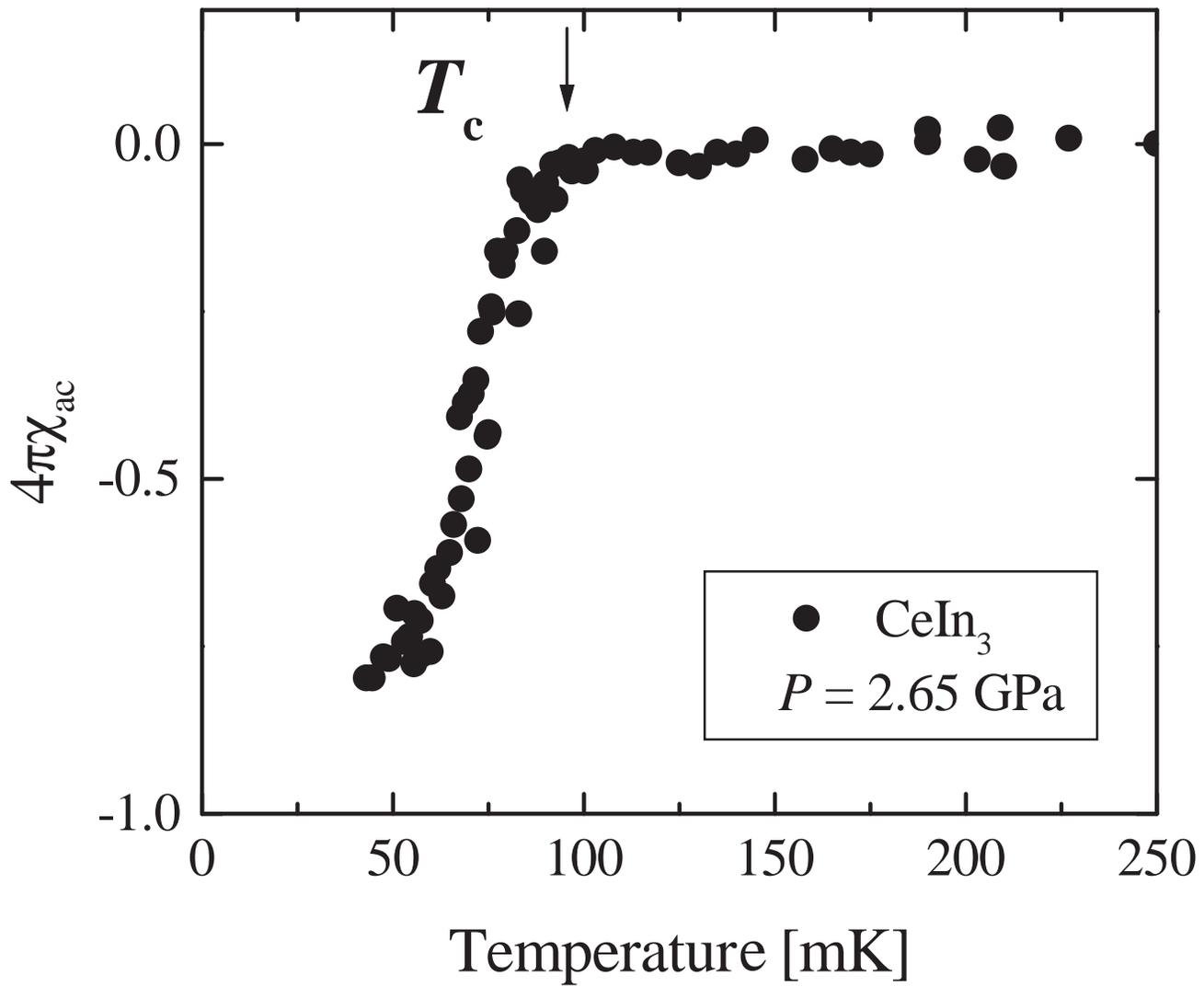
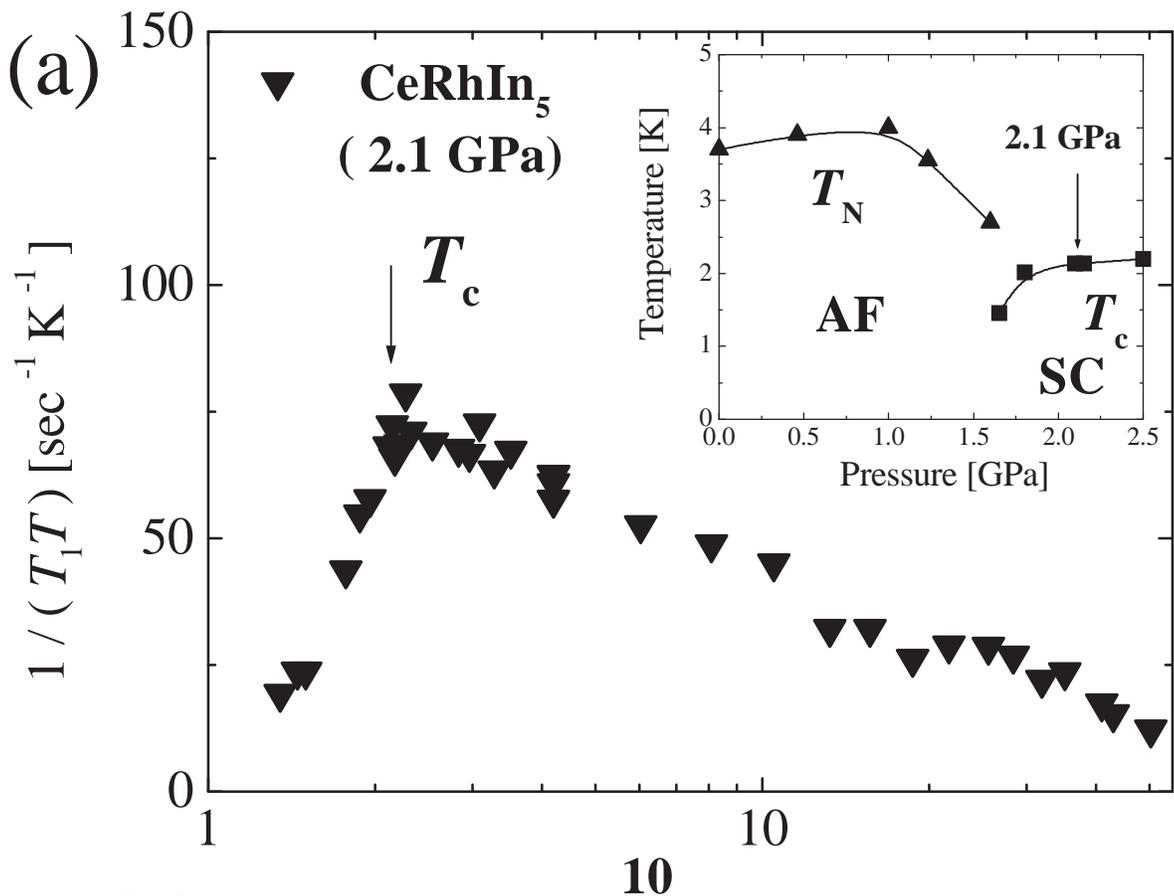
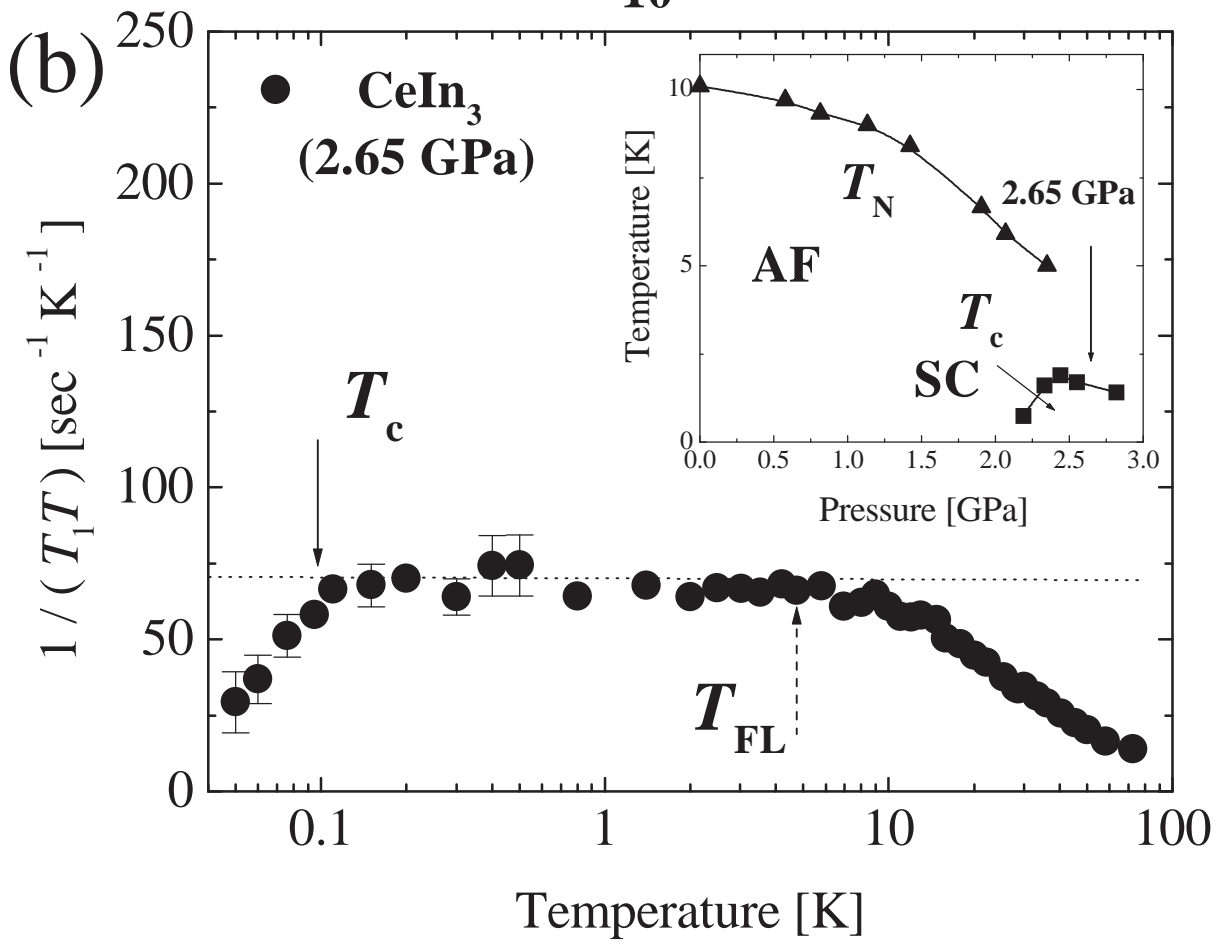
\end{document}